# An Experimental Study of Cryptography Capability using Chained Key Exchange Scheme for Embedded Devices

Mohd Anuar Mat Isa[1], Habibah Hashim[2], Jamalul-lail Ab Manan[3], Syed Farid Syed Adnan[4], Ramlan Mahmod[5]

*Abstract*— After 38 years of birthday Diffie-Hellman Key Exchange (DHKE), there are many proposed improvements in the DHKE protocol to encounter modern security issues. This protocol seems quite simple to be implemented; but it can be vulnerable to many types of attacks. In this work, we propose the Chained Key Exchange scheme as a case study to explore cryptographic computation capability of embedded microcontroller. We choose ARM RaspberryPi board as hardware platform for experimental setup. To enable RasberberryPi *"system on chip"* (SoC) to perform cryptographic computation, we modified the GNU GMP Bignum library to support a simple primitive cryptographic computation in the UBOOT firmware. The main purpose of our study is to determine whether there is any gap between cryptographic protocol/scheme (in term of theoretical) and its engineering implementation. Our scheme will be integrated with Trivial File Transfer Protocol (TFTP) application in the UBOOT firmware. Our proposed scheme in the TFTP protocol will secure the sharing of secrets and symmetric keys (e.g., AES256). After that, the symmetric encryption algorithm can be used to encrypt data in the cases of remote system updates, patching and upgrades (e.g., firmware, kernel or application).

## I. INTRODUCTION

The state of art for key exchange protocol is based on 1976 paper "New Directions in Cryptography" [1], Diffie and Hellman Key Exchange (DHKE) present a secure key agreement protocol that can be carried out over unsecure public communication channels. This protocol seems quite simple to be implemented; but it can be vulnerable to many types of attacks that are based on Number Theory. In this work, we propose the Chained Key Exchange scheme as a case study to explore cryptographic computation capability of embedded microcontroller.

## II. RELATED WORK

After 38 years of birthday DHKE, there are more than 50 proposed improvements in DHKE protocol to encounter security issues such as: J.F. Raymond (2002) [2] collective of attacks in the DHKE protocol and a good idea about how to securely implement the DH protocol in various systems. E. Brickell (2004) [3] provided the first Direct Anonymous Attestation (DAA) scheme based on the strong RSA assumption and decisional Diffie-Hellman assumption. R.C.W Phan (2006) [4] performs cryptanalysis in the DHKE using N-party encrypted different passwords.

E.J. Yoon (2009) [5] proposed an efficient DH's MAC with forward secrecy, key independence and protection against session state reveal attacks. D. Fiore (2010) [6] proposed a new identity based key agreement protocol that is can be implemented over any cyclic group of prime order, where the Diffie-Hellman problem is supposed to be hard. P. Vyas (2012) [7] described various protocols used for key exchange such as freshness of message. H. K. Pathak (2013) [11] proposed two password based of simple three party key exchange protocols via twin Diffie-Hellman problem and showed the proposed protocols provide greater security and efficiency than the existing protocols.

## III. MOTIVATION

Our main motivation in proposing the Chain Key Exchange scheme is to explore the computation capability of embedded microcontrollers such as ARM6 RaspberryPi board in performing cryptographic computation. To explore the possible constraints in the theoretical and experimental designs, we have decided to only use the RaspberryPi board and a USB debug/console cable as experimental setup for the experiment. The RaspberryPi board can support extra I/O functions (add-on card) including sensors, Wi-Fi, camera, sub controllers (e.g., random number generator and customized FPGA with cryptographic functions) and etc. However, we omitted these extra I/O features because we want to study a plain embedded board to perform cryptographic functions.

Chained Key Exchange scheme can be considered as a simple key exchange protocol with only minimal security properties such as forward secrecy and key independence

Manuscript received March 20, 2014; revised April 10, 2014. The authors would like to thank to Ministry of Higher Education (MOHE) for providing the grant 600-RMI/ERGS 5/3 (12/2013), and Universiti Teknologi MARA (UITM) for providing the research grant in this research work.

Faculty of Electrical Engineering, 40450 UiTM Shah Alam, Selangor, Malaysia. [1]anuarls@hotmail.com (corresponding author), [2]habib350@salam.uitm.edu.my, [4]syed_farid@salam.uitm.edu.my
MIMOS Berhad, Technology Park Malaysia, 57000 Kuala Lumpur, Malaysia. [3]jamalul.lail@mimos.my
Faculty of Computer Science & Information Technology, 43400 Universiti Putra Malaysia, Serdang, Selangor, Malaysia. [5]ramlan@fsktm.upm.edu.my

for implementation in the experiment. The main purpose of our study is to determine whether there is any gap between cryptographic protocol/scheme (in term of theoretical) and its engineering implementation. Our intention is to make it as a full working generic security device from an existing microcontroller. From our past experience, we know that is not trivial to do implementation of a cryptographic device which is secured in theoretical, and also secure in its realized device.

*A. Objective*

The objective of this paper is to explore cryptographic computation capability of Chained Key Exchange scheme for an embedded microcontroller in a constrained environment.

*B. Target Application*

This study will attempt to establish a secure and trust based key exchange protocol in the embedded controller. The term of "trust" is based on our previous work in Trusted Computing wherein *"How can we be assured device(s) and system(s) are trusted if we use trusted computing platform (e.g., TPM) as root of trust?"* [8]. For this experiment, we do not use Trusted Platform Modules (TPM), but rather, we explore the concept of *"chain of trust"* in the cryptographic scheme, i.e. chain of trust of secret keys. The *"chain of trust of secret keys"* allows our protocol to verify that new communication with third parties is the same as previous communication through secure transitive sessions. The proposed protocol would be useful for lightweight or smart embedded device to identify whether an adversary is trying to intrude into the confidential communication. Energy usage becomes major factor for operational consideration by lightweight devices especially for deployment without compromising on security. In our proposal, we will use minimal I/O peripheral to reduce energy consumption, and at the same time yield high cryptographic computation in the autonomous environment.

In the long term run, our scheme will provide an implementation of Secure Trivial File Transfer Protocol (TFTP) application in the UBOOT firmware. It will ensure remote system updates and patching (e.g., firmware, kernel or application) processes are secure from attacks which aim to eavesdrop and modify the TFTP packet. The target employment of Secure TFTP protocol is in the Wi-Fi Access Points, remote base stations, wireless sensor nodes and etc.

## IV. EXPERIMENTAL SETUP

*A. Chained Key Exchange Scheme*

**A1. Initialization of pre-shared knowledge between two parties** (assume that their names are Along and Busu). This pre-shared knowledge must happen during production, physical exchange or through a trusted communication.

*Along generates*: core root of trust (crt) of random numbers $a_{crt}, p_{crt}, g_{crt} \in \mathbb{Z}_n^*$. Along stores $a_{crt}, p_{crt}, g_{crt}$ in the non-volatile memory equipped with physical tamper-resistant technology. Along's secret $a_{crt}$ is protected using user authentication. Then, Along computes:

$$A_{crt} \equiv g_{crt}^{a_{crt}} \pmod{p_{crt}} \quad \text{where } A_{crt} \text{ size of } n$$

Along share public information $A_{crt}, p_{crt}, g_{crt}$ with Busu.

*Busu generates*: core root of trust (crt) of random numbers, length of n-bits $b_{crt} \in \mathbb{Z}_n^*$. Busu stores $b_{crt}, p_{crt}, g_{crt}$ in the non-volatile memory equipped with physical tamper-resistant technology. Busu's secret $b_{crt}$ is protected using user authentication. Then, Busu computes:

$$B_{crt} \equiv g_{crt}^{b_{crt}} \pmod{p_{crt}} \quad \text{where } B_{crt} \text{ size of } n$$
$$Key_{crt} \equiv A_{crt}^{b_{crt}} \pmod{p_{crt}} \quad \text{where } Key_{crt} \text{ size of } n$$

Busu share public information $B_{crt}$ with Along.

*Along computes*:

$$Key_{crt} \equiv B_{crt}^{a_{crt}} \pmod{p_{crt}} \quad \text{where } Key_{crt} \text{ size of } n$$

Finally, both parties store the matching secret $Key_{crt}$ in the non-volatile memory with physical tamper resistant technology wherein the secret key is protected using user authentication. This initialization of crt key is less likely to be computed compared to the session key and chain key. We assumed $Key_{crt}$ computation happens only in safe environments (e.g., during production of embedded device) and no integrity verification of the messages is required. Furthermore, an adversary would not be able to eavesdrop this information because it happens in close environments. This initialization scheme has been originated from DHKE [1] scheme.

**A2. Initialization of "chained of session key" between two parties.**

In this scenario, the "chain of session key" occurs in open communication channel; hence it is still vulnerable to adversaries.

*Along generates*: chain of session of random numbers.
$a_i, p_i, g_i \in \mathbb{Z}_n^*$ and i = 0, where 0 is initial chain sequence and
$a_i \neq Key_{crt}$, $p_i \neq p_{crt}$, $g_i \neq g_{crt}$
Along stores $a_i, p_i, g_i$ in the non-volatile memory equipped with physical tamper-resistant technology. Along's secret $a_{crt}$ is protected using user session authentication. Then,
*Along computes:*

$$A_i \equiv g_i^{a_i} \pmod{p_i} \quad \text{where } A_i \text{ size of } n$$

Along share public information $A_i, p_i, g_i$ with Busu.

*Busu generates*: secret $b_i$ of random numbers, of length n-bits.
$b_i \in \mathbb{Z}_n^*$ and i = 0, where 0 is initial chain sequence
Busu stores $b_i, p_i, g_i$ in the non-volatile memory equipped with physical tamper-resistant. Busu's secret $a_i$ is protected using user session authentication. Then, *Busu computes*:

$$B_i \equiv g_i^{b_i} \pmod{p_i} \quad \text{where } B_i \text{ size of } n$$
$$Key_i \equiv A_i^{b_i} g_i^{Key_{crt}} \pmod{p_i} \quad \text{where } Key_i \text{ size of } n \quad (1)$$

Busu share public information $B_i$ with Along.

*Along computes*:
$$Key_i \equiv B_i^{a_i} g_i^{Key_{crt}} \pmod{p_i} \quad \text{where } Key_i \text{ size of } n \quad (2)$$

Session $Key_{i=0}$ needs to be verified before being stored or used (explained in Section **A3** for details). The shared secret $Key_i$ requires 3n length of exponential computation or $O(3n)$ as shown in (3). For the next session of key computation, we use key derivative function to derive $\overline{Key_{i=0}}$ from $Key_{i=0}$. Therefore, we conclude that the $Key_{i=1}$ as follow:

$$Key_1 \equiv g_1^{a_1} \cdot g_1^{b_1} \cdot g_1^{\overline{Key_0}} \pmod{p_1} \quad (3)$$

However, there is no guarantee that $Key_i$ in (1-3) will be the size of n after the successful key exchange process. In worst case scenario, it produces a weak key with a short length.

Therefore, $Key_i$ needs to be checked[1]/discarded before we can proceed to the verification process as it will be explained next.

*A3. Verification of "chained of session key"*

The purpose of verification is to ensure that both parties will synchronize and Message Authentication Code (MAC) the generated $Key_i$ from Section *A2*. If successful, both parties will store the matching secret $Key_i$ in the non-volatile memory equipped with physical tamper resistant technology wherein the secret key is protected using user authentication. The previous $Key_{i-1}$ (if it still exists) is safely wiped out from nonvolatile memory. We define nonce as $nonce_i \in (\mathbb{Z}_n^*)$ and initial with $nonce_{i=0} = 0$'s string of length n.

*Along performs hashing function*:
$if\ i = 0;\ digest_{Along_i} = hash(Key_i \parallel Key_{crt} \parallel nonce_i)\ else$
$if\ i > 0;\ digest_{Along_i} = hash(Key_i \parallel \overline{Key_{i-1}} \parallel digest_{i-1} \parallel nonce_i)$

Along expects $digest_{Busu_i}$
$if\ i = 0;\ digest_{Expected\ Busu_i} = hash(Key_{crt} \parallel Key_i \parallel nonce_i)\ else$
$if\ i > 0;\ digest_{Expected\ Busu_i} = hash(\overline{Key_{i-1}} \parallel Key_i \parallel digest_{i-1} \parallel nonce_i)$

Along shares public information $digest_{Along_i}$ with Busu.

*Busu performs hashing function*:
$if\ i = 0;\ digest_{Busu_i} = hash(Key_{crt} \parallel Key_i \parallel nonce_i)\ else$
$if\ i > 0;\ digest_{Busu_i} = hash(\overline{Key_{i-1}} \parallel Key_i \parallel digest_{i-1} \parallel nonce_i)$

Busu expects $digest_{Along_i}$:
$if\ i = 0;\ digest_{Expected\ Along_i} = hash(Key_i \parallel Key_{crt} \parallel nonce_i)\ else$
$if\ i > 0;\ digest_{Expected\ Along_i} = hash(Key_i \parallel \overline{Key_{i-1}} \parallel digest_{i-1} \parallel nonce_i)$

Busu shares public information $digest_{Busu_i}$ with Along.

*Along verifies*:
$digest_{Expected\ Busu_i} = digest_{Busu_i}$

*Busu verifies*:
$digest_{Expected\ Along_i} = digest_{Along_i}$

Observe that in hashing function between Along and Busu, the sequence of hashing function is different in the first parameter and the second parameter for the hashing input. This will guarantee that the hashing digests of Along and Busu are different for the MAC authentication process. To protect from an attack based on Number Theory, such as "degenerate message attack" [2], we need to ensure random secrets, public parameters and $Key_i$ are not recycled numbers. For the next session, we must use a secure one way key derivation function to derive $\overline{Key_i}$ from $Key_i$ to avoid using previous key.

After that, both parties store $Key_i$ that has been successfully verified. In case of failure, the digest need to be retransmitted (retry) because errors may happen in communication medium when using non-reliable network. Failure to do correction and verification within the allowed number of retries, the verification process is considered invalid and the chain of session $i$ must be dropped. All temporary data in Section *A2* must also be safely wiped out from volatile memory. If this problem happens, we can consider that there are problems i) in the communication channel, ii) an active adversary is impersonating either parties or iii) an active adversary has tampered the digest.

*B. Embedded System*

We decided to use RasberberryPi Model B (Fig. 1) with specifications: BCM2835 (ARMv6k) 700 MHz, 512MB RAM, 16GB SD memory card, 10/100 Ethernet port. This board is widely used for system prototyping or experiment, system controller, surveillance system, cluster nodes, embedded programming etc. We have done literature review on past works and we found that it is not well explored yet. From here, we decided to conduct cryptographic primitive computation using this board. Among the major issues need to be considered when using this board are GCC ARM compiler and GMP Bignum [9] library to compute numbers beyond 32-bit integers (e.g., exponential, modular, etc.).

We conducted an experiment based on "one-group pretest-posttest" [10][11] experimental design to evaluate performance measurement of Chained Key Exchange scheme. The first test group was conducted in application layer (user space) through Linux Raspbian "wheezy" Kernel using precompiled image "2013-07-26-wheezy-raspbian.zip" [12]. The second test group was conducted in firmware layer (bare metal) using Denx U-Boot [13] as platform for bare metal execution of our scheme. U-Boot provides cross platform execution because it supports multiple embedded architecture such as ARM, MIPS, PPC, x86, 68k, Nios and etc. Therefore, we are confident that with a very minimal configuration, our protocol can be deployed in multiple embedded systems.

To enable RasberberryPi "system on chip" (SoC) to perform cryptographic computation, we modified the GMP Bignum version gmp-5.1.0 [9] library for a simple primitive cryptographic library. However, major modification is required in a bare metal system because of missing C library and its dependencies in the U-Boot. We noted that most of standard C libraries are meant for application and kernel layers, but not in firmware layer. This means that most of C libraries in firmware programming are minimal for the purpose of startup for the device and loading an operating system kernel for a system to boot up. To reduce the complexity, we modified the "mini-gmp" section to diminish the dependency problems. The modified "mini-gmp" is encoded in the first and second group experiments for fairness of execution and timing measurements. Our new "mini-gmp" library support the major functions for cryptographic computations such as mpz_init(), mpz_clear(), mpz_t, mpz_set_str(), mpz_powm(), mpz_get_str(), mpz_cmp(), mpz_sub(), mpz_add(), mpz_ui_pow_ui(), mpz_gcdext(), mpz_invert() clock(), SHA512(), and etc. Based on our previous work, we work on the communication protocol for two sets of RasberberryPi board using a secure TFTP protocol for smart environment [14]. This work [14] we discussed the modification of U-Boot's TFTP protocol to support a secure key exchange and data encryption.

---

[1] We can use generated key with a key length less than n (e.g. $(n-2)$ length); but we need to use a secure one way key expander/derivation function to fill-up (or padding) the less significant part of number in $(n-2)$ length. However this is very risky when the $(n - \left(\frac{n}{2}\right))$ length is too short.

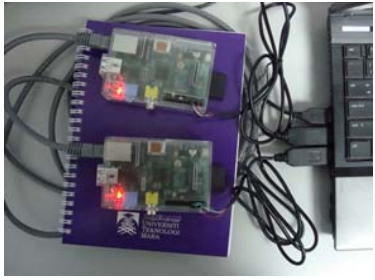

**Fig. 1.** Experimental Testbed.

## V. SECURITY ANALYSIS

### A. Protocol Security

The underlying security principles for Chain Key Exchange scheme are based on the i) *Decision Diffie-Hellman Problem (DDHP)* [15] and ii) *one way hash function* [16]. To solve DDLP, an adversary needs to find $a_i, b_i, a_{crt}, b_{crt}$ in Section A1. For example, $g_i$ is a primitive root of $p_i$ of order $n$.

$$Key_1 \equiv g_i^{a_i} \cdot g_i^{b_i} \cdot g_i^{\overline{Key_i}} \pmod{p_1} \text{ if } n > 2^{80}$$

It is not feasible to use brute force to find $Key_1$ in above equation. The fast algorithm to do brute force for modular exponential requires time complexity of $O(2^n)$ where $n$ is a number of steps in bits. We use hash function in Section *A3* for verification of secret key that is generated during chain of key exchange. We consider a simple Message Authentication Code (MAC) hash function to verify Along's and Busu's shared secret key in the communication. We also assume that the one way hash function is a secure pseudo random generator [16] wherein the adversary cannot distinguish the hashing output (digest) from input to the hashing algorithm. For the moment, we use a standard SHA512 because it has been assumed to be secure[2] based on many recent literature [17][18][19][20][21]. Based on these two underlying security assumptions stated above, we consider our protocol as secure against the following outsider attacks:

i. Session state reveal attack: The Chain Key Exchange scheme is secure even when an adversary manages to know the current session secrets $a_i, b_i$ in Sections *A1* and *A2* above. An adversary cannot compute $Key_0 \equiv g_0^{a_0 \cdot b_0 \cdot Key_{crt}} \pmod{p_0}$ or $Key_i \equiv g_i^{a_i \cdot b_i \cdot \overline{Key_{i-1}}} \pmod{p_i}$ because root key $Key_{crt}$ and chain key $\overline{Key_{i-1}}$ are unknown.

ii. Forward secrecy: The Chain Key Exchange scheme is still secure even when an adversary manages to know the previous session secrets $a_{i-1}, b_{i-1}, Key_{i-1}, Key_{crt}$ in Sections *A1*. An adversary cannot compute $Key_i \equiv g_i^{a_i \cdot b_i \cdot Key_{crt}} \pmod{p_i}$ or $Key_i \equiv g_i^{a_i \cdot b_i \cdot \overline{Key_{i-1}}} \pmod{p_i}$ because current session secrets $a_i, b_i$ are unknown.

iii. Key independence: Based on the forward secrecy, the scheme is secure when an adversary manages to know the previous secret key $Key_{i-1}$ because of a new session secrets $a_i, b_i, Key_i$ are generated and computed independently from all previous sessions. An adversary cannot attack a new chain of session using previous knowledge; even if those previous session keys are already broken and exposed. This requires the total length of all secrets ($a_i, b_i, \overline{Key_{i-1}}$) > $3n$ length of bits and each secret $\geq n$ length of bits. At present, it is acceptable to use $2^n$ where $n = 1024$ bits length. Based on DDHP assumption, an adversary knowledge of $Key_{i-1}$ is considered infeasible to derive $Key_i \equiv g_i^{a_i \cdot b_i \cdot \overline{Key_{i-1}}} \pmod{p_i}$ because of unknown secrets $a_i, b_i$ (total $2^n$ of secrets length is considered infeasible to attack). Therefore, using new ephemeral random numbers $a_i, b_i$ with appropriate length, our scheme is resistant against "Denning-Sacco" [22] [7] attack.

iv. Key derivation function attack: Our scheme can resist the Burmester triangle attack [23] because we include 1) key derivate function to derive $\overline{Key_i}$ from $Key_i$ and 2) the public parameters $A_i$ and $B_i$ only consist 2n length of exponential computations of 2 secrets (such as $a_i, b_i$) in Section **A2**. Both public parameters could never be computed with $g_i^{Key_{crt}}$ or $g_i^{\overline{Key_{i-1}}}$ (due to the missing the 3rd secret value in public parameters as). An adversary can impersonate Along when communicated with Busu in (4-6) and vice versa (7-9). Let us assume that an adversary can also manages to break previous session $i-1$ with adversary's knowledge of the secrets $a_i, b_i$ and $Key_i$. However, the adversary will fail to break (11) of our protocol because of the two reasons mentioned above. With a strong key derivation function, we can secure future session keys because it destroys algebraic relationships[3] in between old session keys and new session keys.

$$A_{adv} \equiv g_i^{a_{adv}} \pmod{p_i} \quad (4)$$

$$B_{busu} \equiv g_i^{b_{busu}} \pmod{p_i} \quad (5)$$

$$Key_{adv+busu} \equiv g_i^{a_{adv} \cdot b_{busu}} \pmod{p_i} \quad (6)$$

$$A_{along} \equiv g_i^{a_{along}} \pmod{p_i} \quad (7)$$

$$B_{adv} \equiv g_i^{b_{adv}} \pmod{p_i} \quad (8)$$

$$Key_{along+adv} \equiv g_i^{b_{adv} \cdot a_{along}} \pmod{p_i} \quad (9)$$

(6) and (9) $Key_{along+adv} + Key_{adv+busu}$

$Key_{along+busu} \equiv$

$$g_i^{a_{adv} \cdot b_{busu}} + g_i^{b_{adv} \cdot a_{along}} + g_i^{Key_{i-1}} - (g_i^{a_{adv}} + g_i^{b_{adv}})(mod\ p_i) \quad (10)$$

From (10)
$$Key_{along+busu} \equiv g_i^{a_{along} \cdot b_{busu} \cdot Key_{i-1}} \pmod{p_i} \quad (11)$$

---

[2] We choose to use SHA-512 because it receives wider input block in size of 1024 bits and output hashing digest in size of 64 words or strings. With this size of input block, it can be pushed all at once for n = 1024 bits. It helps to produce a better MAC in terms of 64 words of hashing digest.

[3] We considered that the key derivation function is a secure one way pseudo random function wherein both sides (Along and Busu) are using the same function to derive $\overline{Key_i}$ from $Key_i$.

Key agreement and confirmation: Authentication based on hashing MAC is used to protect message integrity, replay attack, man-in-the-middle attack and chain of session keys in Section *A3*.

### B. Hardware Security

Physical access to RaspberryPi board is a major problem in the product deployment. An attacker can easily remove the SD card memory from the board and use memory card reader to access all data in the SD card. We can encrypt the memory region in SD card using common techniques such as disk encryption[4]. At the moment, there is no temper resistant circuit in the devices, it can be easily broken if an attacker has physical access to the device. We need extra I/O devices that are connected to board with implementation of physical temper resistant (e.g., memory), physically unclonable function (PUF), and etc.

### C. Security Checkpoints for Implementation

We have listed below some security concerns and guidelines for the implementation and deployment. For additional reading, one may refer to [2] for further understanding of good security implementations.

i. Weak primes group and order: Vulnerability to Pohlig-Hellman algorithm, composite order subgroup for $p = 2q + 1$, Pollard Lambda algorithm, Number Field Sieve algorithm and etc.
ii. Weak randomness of number generator and randomness seeds.
iii. Key management and memory segregation for many sessions for different applications.
iv. Choosing k size of symmetric key encryption, we need 2k size of exponential of each secrets to secure against Pollard's rho method (finding collisions between values computed in a large number space)
v. Never recycle any keys after usage.
vi. Generate a new session key for symmetric encryption before it reaches a point that is encrypted message or hashing MAC will produce collisions.
vii. There is possibility that the proposed scheme is vulnerable to timing attacks because of exponential computation in RaspberryPi. If an adversary can effectively and precisely determine the computation of 700 MHz CPU to compute public parameters from Along's or Busu's secrets. However, an implementation of key derivate function in our scheme will render the timing issue.
viii. Check data type of numbers (int, unsigned, sign, strings, byte stream and etc), valid values range or size of received in the I/O buffer before loading it into cryptographic functions.

## VI. RESULT & DISCUSSION

Table I shows the result of experiments wherein each experiment sessions are measured in seconds. All experimental setup are using the same codes and input data such as primes number, ephemeral secrets and etc. We use pre-generated random numbers and other input data to guarantee that is each experimental setup are free from bias of randomness when using random number generator. The most fascinating and unexpected finding is that the performance of the protocol in bare metal is slower than execution in the operating system[5]. The results shown the computation time to exchange one secret key based on Sections *A2, A3* and network performance. We omitted the Section *A1* because of it can be generated using original DHKE or pre-install the device with core-root-key during manufacturing process. Based on result in the Table 1, it is required 12~14 computation times to exchange a secret key in length of 2048 bits to compare with 1024 bits secret key. We conclude that is our experiment meets the minimal target objective in the research and development of Secure TFTP application in the UBOOT. It is extremely difficult and unrealistic to make a 360 degrees security fortification in order to fulfill all security requirements such as side-channel security and [2] because of resource limitation in the embedded devices.

TABLE I
An Average of Chain of Key Exchange Scheme (One Chain Cycle) Performance in Second.

| No | Experimental Setup | $n$=1024 Bits Key | $n$=2048 Bits Key |
|---|---|---|---|
| 1 | RaspberryPi, Bare Metal, default UBOOT setting (disable CPU's internal caches), modified GMP Bignum library. | 8.364 | 63.998 |
| 2 | RaspberryPi, Bare Metal, modified UBOOT codes & setting (enable CPU's internal caches), modified GMP Bignum library. | 1.887 | 14.671 |
| 3 | RaspberryPi, console, *2013-07-26-wheezy-raspian* OS image, modified GMP Bignum library. | 1.571 | 12.035 |
| 4 | RaspberryPi, GUI console (startx), *2013-07-26-wheezy-raspian* OS image, modified GMP Bignum library. | 1.607 | 12.093 |
| 5 | HP Elitebook 8440w, 2.8 GHz i7, 8 GB RAM, Debian 6 OS and GUI (startx). | 0.196 | 1.479 |

## VII. SUGGESTIONS AND FUTURE WORK

Varied research and developments works must be done to improve the proposed scheme. For the security protocol, we plan to investigate the usage of Diophantine Equation Hard Problem (DEHP) [24] [25] to provide faster cryptographic computation to replace the DDHP because of it is slower in the exponential computation. We want to include a zero energy consumption of true random number generator using a wide band passive antenna and frequency hopping in receiver (Rx) antenna. This wide band receiver will consume analog signals or frequencies in the air space as input for powerless random number generator (or green RNG). For a current work, our research group is focusing in developing a secure TFTP communication in radio frequency (RF) using CISECO B023 Slice of Radio RF transceiver [26]. For the cryptographic computing, we want to proof our hypothesis that executing in the bare metal is faster than executes in the generic operating system. This required additional programming in the UBOOT to improve CPU's caches and external caches, DMA and pipelining processes in the MMU.

---

[4] The size of UBOOT firmware with implementation of our scheme is 185,584 bytes. We can add user authentication module in the UBOOT firmware when user want to access our scheme (encrypted memory regions of secrets and keys). We considered it as a future work.

[5] Before we conduct the bare metal experiment, we used to believed that execution in the bare metal is faster because it is free from any disturbance of operating system [27] scheduling or context switching of processes (our hypothesis). Result of Table 1, prove that we need to consider the hardware implementation to make it works faster. In the bare metal environment; there is no pipelining and other OS's caching techniques that helps to improve the performance of data fetching and CPU's instruction fetching for CPU's execution process.

## VIII. CONCLUSION

We presented our Chain of Key Exchange scheme with experimental results. We succeed to fulfill our research objective and that is to explore cryptographic computation capability of embedded microcontroller in the constrained environment. Our major contribution is the integration of our scheme for secure TFTP application in the UBOOT firmware. However, the result in the experiments is not pretty as our hypothesis expectation. The result shown that is the performance of our scheme in the bare metal is slower compared to execution in the operating system. At the moment, we considered this is the most flopped research hypothesis for this research project. One might think that is our scheme and experiment are failure work. But, we will keep trying with new ideas and improve it even that are our $100^{th}$ flopped ideas and *God rewards fool*[6].

---

[6] *"…the way to get to the top of the heap in terms of developing original research is to be a fool, because only fools keep trying. You have idea number 1, you get excited, and it flops. Then you have idea number 2, you get excited, and it flops. Then you have idea number 99, you get excited, and it flops. Only a fool would be excited by the 100th idea, but it might take 100 ideas before one really pays off. Unless you're foolish enough to be continually excited, you won't have the motivation, you won't have the energy to carry it through. God rewards fools."* - Whitfield Diffie and Martin Hellman, in The Code Book, p. 256.